\documentclass[]{spie}  %>>> use for US letter paper
\usepackage{amsmath,amsfonts,amssymb}
\usepackage{graphicx}
\usepackage[colorlinks=true, allcolors=blue]{hyperref}

\title{ LOCNES: a solar telescope to study stellar activity in the near infrared}

\author[a]{Claudi R.}
\author[b]{Ghedina A.}
\author[c]{Pace E.}
\author[d]{Di Giorgio A.M.}
\author[a]{D'Orazi V.}
\author[c]{Gallorini L.}
\author[f]{Lanza A.F.}
\author[d]{Liu S.J.}
\author[e]{Rainer M.}
\author[e]{Tozzi A.}
\author[a]{Carleo I.}
\author[g]{Maldonado Prado J.}
\author[g]{Micela G.}
\author[h]{Molinari E.}
\author[b]{Poretti E.}
\author[i]{Phillips D.}
\author[j]{Tripodo G.}
\author[b]{Cecconi M.}
\author[b]{Galli A.}
\author[b]{Gonzalez M. D.}
\author[b]{Guerra Padilla V.}
\author[b]{Guerra Ram\'on J.G.}
\author[b]{Harutyunyan A.}
\author[b]{Hern\'andez C\'aceres N.}
\author[b]{Hern\'andez D\'iaz M.}
\author[b]{Lodi M.}
\author[b]{P\'erez Ventura H.}
\author[b]{Riverol Rodr\'iguez A. L.}
\author[b]{Riverol Rodr\'iguez C. A.}
\author[b]{San Juan G\'omez J.}

\affil[a]{INAF Astronomical Observatory of Padova, vicolo Osservatorio, 5, Padova, Italy}
\affil[b]{INAF -- Fundaci\'on Galileo Galilei, Rambla Jos\'e Ana Fern\'andez P\'erez, 7,  Bre$\tilde{n}$a Baja--TF, Spain}
\affil[c]{Dep. of Physics and Astronomy, Universit\'a degli studi di Firenze, Firenze, Italy}
\affil[d]{INAF-- IAPS, via del Fosso del Cavaliere, 100, Roma, Italy}
\affil[e]{INAF--Astrophysical Observatory of Arcetri, Largo Enrico Fermi,5, Firenze, Italy}
\affil[f]{INAF--Astrophysical Observatory of Catania, Via S.Sofia 78, Catania, Italy}
\affil[g]{INAF--Astronomical Observatory of Palermo,P.zza del parlamento, 1, Palermo, Italy}
\affil[h]{INAF--Astronomical Observatory of Cagliari,Via della Scienza, 5, 09047 Cuccuru Angius, Selargius (CA), Italy}
\affil[i]{Harvard-Smithsonian Center for Astrophysics, Cambridge, MA USA 02138}
\affil[j]{Universit\'a di Palermo, Scuola delle Scienze di base e applicate, Dip. di Fisica e Chimica, Piazza Marina, 61, 90133, Palermo (Italy)}

\authorinfo{Further author information: (Send correspondence to R.Claudi)\\R. Claudi: E-mail: riccardo.claudi@inaf.it}

\begin{document} 
\maketitle

\begin{abstract}
LOCNES (LOw-Cost NIR Extended Solar telescope) is a solar telescope installed at the TNG (Telescopio Nazionale Galileo). It feeds the light of the Sun into the NIR spectrograph GIANO-B through a 40-m patch of optical fibers.
LOCNES has been designed to obtain high signal-to-noise ratio spectra of the Sun as a star with an accurate wavelength calibration through molecular-band cells. This is an entirely new area of investigation that will provide timely results to improve the search of telluric planets with NIR spectrographs such as iSHELL, CARMENES, and GIANO-B. We will extract several disc-integrated activity indicators and average magnetic field measurements for the Sun in the NIR. Eventually, they will be correlated with both the RV of the Sun-as-a -star and the resolved images of the solar disc in visible and NIR. Such an approach will allow for a better understanding of the origin of activity-induced RV variations in the two spectral domains and will help in improving the techniques for their corrections. 
In this paper, we outline the science drivers for the LOCNES project and its first commissioning results. 
\end{abstract}


\section{INTRODUCTION}
\label{sec:intro}  % \label{} allows reference to this section
The search for very small extrasolar planets with the Radial Velocity technique is plagued by the stellar jitter due to the activity of the star, because stellar surface inhomogeneities including spots, plages and granules, induce perturbation that hide the planetary signal. This kind of noise is poorly understood in all the stars but the Sun due to their unresolved surfaces. For these reasons the effect of the surface inhomogeneities on the measurement of the Radial velocities are very difficult to characterize. On the other hand, a better understanding of these phenomena can allow us to made a step further in our knowledge of solar and stellar physics. The latter will allow to acquire more skills in the art of developing optimal correction techniques to extract true stellar radial velocities. A viable way to tackle these problems is to observe the Sun as a star getting time series of radial velocities in order to disentangle the several contribution of the stellar (solar) RV jitter\cite{dumusqueetal2015apjl,marchwinskietal2015,haywoodetal2016}. 

To obtain long-term observation of the Sun as a star with state-of-the-art sensitivity to RV changes, since July 2015, a Low-Cost Solar Telescope (LCST) has been installed\cite{dumusqueetal2015apjl, phillipsetal2016} on the outside of the dome at TNG to feed solar light to the HARPS-N spectrograph (0.38-0.69um; R=115000).  Following the example of LCST, several spectrographs both in the north and south hemispheres have been equipped with such solar telescopes. Examples\cite{dumusqueetal2020arxiv} are HARPS at 3.6m of ESO, EXPRES at the Discovery Channel Telescope, NEID at the KNPO, and MAROON-X at Gemini north Telescope. They are all spectrographs dedicated to the measurement of high precision radial velocity in the visible wavelength range. The use of LCST allowed HARPS-N to collect about three years of Solar spectra with a cadence as high as 5 minutes. This huge mole of data on solar radial velocities and activity-index indicators has been analyzed by several investigators \cite{colliercameron2019mnras,milbourneetal2019apj, maldonadoetal2019aa,miklosetal2020apj,langellieretal2020arxiv}.

In the last decade, new near-infrared (NIR) spectrographs have been built that can measure high precision RVs and thanks to the advent of better technology, specifically in calibration, these spectrographs can approach the precision that routinely is obtained in the visible. Among these NIR spectrographs, there is the high-resolution NIR spectrograph of the Telescopio Nazionale Galileo (TNG) named GIANO--B. The Italian exoplanetary community through the {\it Progetto Premiale WOW} funded the refurbishment of GIANO-A\cite{olivaetal2012} in the framework of the GIARPS project\cite{claudietal2017, claudietal2018spie}.
We planned to replicate the optomechanical configuration of the LCST extending the wavelength range to the NIR feeding GIANO--B (0.9 to 2.5um).  In this paper, we want to describe the science drivers of the project, the status of LOCNES, its operations, and its \textit{first light} data.

\section{Science drivers}
\label{sec:sciencedrivers}  % \label{} allows reference to this section
%High Resolution spectrographs are used for high precision radial velocity measurements in the search for extrasolar planets evaluating the Doppler shift of the spectral lines of the star due to the presence of a low mass companion. In this task, these instruments become more and more precise also because people apply the two techniques of the simultaneous thorium and absorbing cell. 
The precision in the measurement of RVs reached such a low limit (the best precisions obtained nowadays are of the order of 0.3 m/s rms) 
that the detection of very small planets (or planets in wider orbits) has to tackle with low amplitude stellar activity. In fact, besides the instrumental noise, the radial velocity technique is also hampered by astrophysical sources of noise mainly due to the host star itself like, for example, the stellar pulsations, surface granulation, and stellar jitter caused by star spots or other instabilities in the stellar atmosphere. At the 1 m/s level of precision, in fact, these physical phenomena in stellar photospheres give significant signals that can hide planetary radial-velocity signatures if not properly modeled. Solar-type stars have an outer convective envelope that exhibits variability on different timescales. 
%There are several sources of noise that are necessary to take into account in high precision radial velocity measurements. 
In any case, the signal of a planet and that due to activity should be distinguishable because of the different physical origin. But, so far, this is not the case. 
Understanding the RV signatures of stellar activity, especially those at the stellar rotation time-scale, is essential to better understand the short and very short time scale variability of stars, but also to improve our ability to detect and characterize (super-)Earths and even small Neptunes in orbits of a few days to weeks. 

All the sources of these stellar noise are inherent to stellar surface inhomogeneity, but, up to now it was only possible to obtain precise, frequent, full- disk RV measurements derived from the full optical spectrum for stars other than the Sun. However, the surfaces of other stars are not resolvable; and therefore all the information used to study surface inhomogeneities is indirect, including: photometric flux (e.g., \citenum{boisseetal2009,aigrainetal2012}), bisector of spectral lines or of the cross correlation function depending on the activity level of the star (e.g., \citenum{vogt1987,quelozetal2001,figueiraetal2013,dumusqueetal2014}), or the calcium chromospheric activity index \cite{noyesetal1984}. Because it is extremely difficult to infer from such indirect observables the size, location, and contrast of surface inhomogeneities, understanding in detail the RV perturbations produced by these features is a major challenge. 
This is true in particular today that RV searches started to target young and active stars. 

The Sun is the only star whose surface can be directly resolved at high resolution, and therefore constitutes an excellent test case to explore the physical origin of stellar radial -- velocity (RV) variability. The contemporaneous observation of the Sun as a star in both the VIS and in the NIR will give  us the possibility to better understand the effect of surface inhomogeneities on RVs by obtaining precise full-disk RV measurements of the Sun. This approach of observing the Sun as a star allows us to directly correlate any change in surface inhomogeneities observed by solar satellites like the Solar Dynamics Observatory (SDO \cite{pesnelletal2012}) with variations in the full-disk RV.

   \begin{figure} [ht]
   \begin{center}
   \begin{tabular}{c} %% tabular useful for creating an array of images 
   \includegraphics[height=8cm]{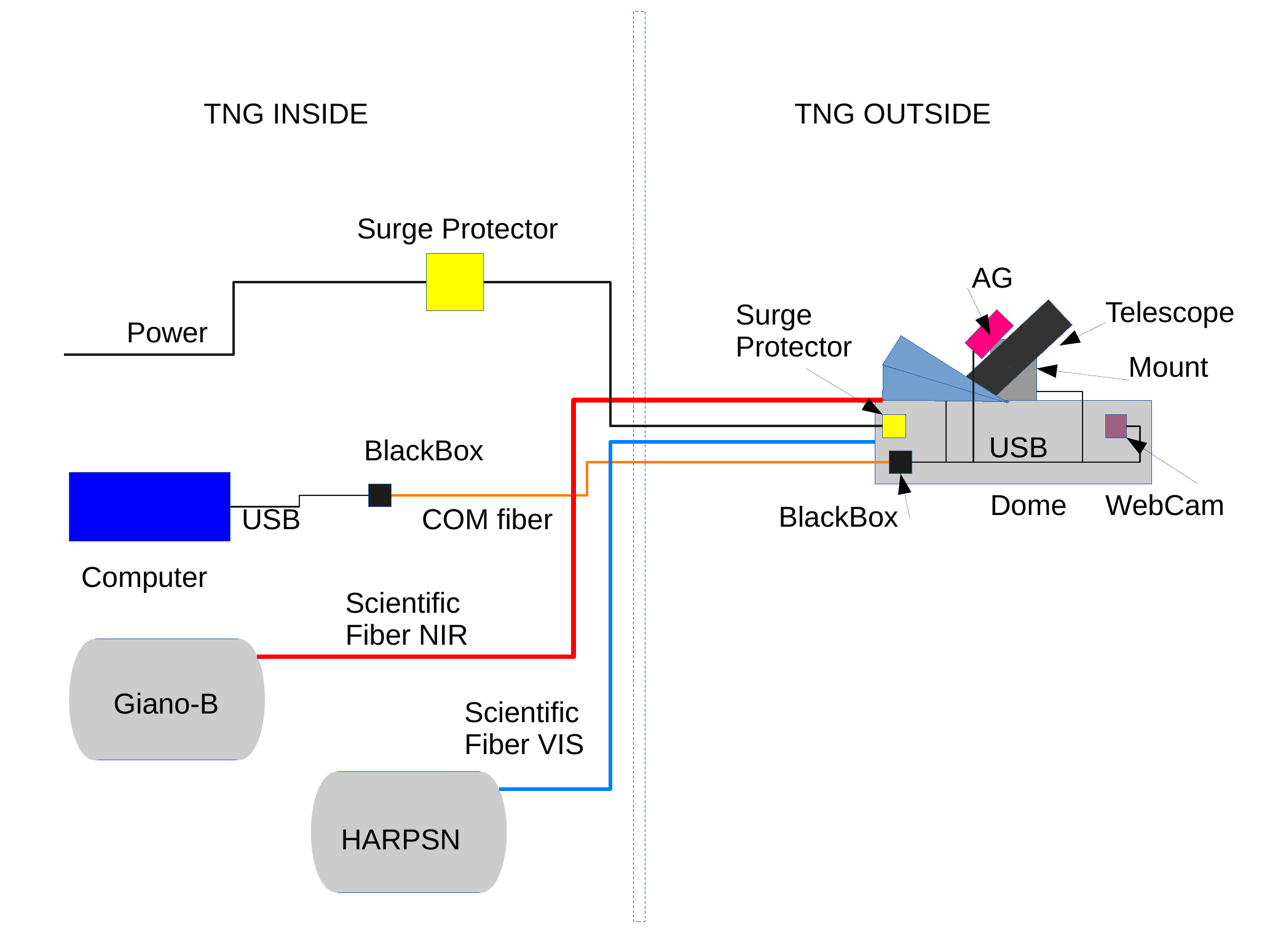}
   \end{tabular}
   \end{center}
   \caption[example] 
%>>>> use \label inside caption to get Fig. number with \ref{}
   { \label{fig:scheme} 
LOCNES system scheme. All main part are indicated together with the connections to the control PC and spectrographs inside the TNG structure.}
   \end{figure} 

\subsection{The K Band Justification}
\label{ssec:kband}  % \label{} allows reference to this section
The measurements of the sun-as-a-star radial velocity was performed by Deming \& Plymate\cite{demingandplymate1994} by means of the molecular bands of CO around 2.3 microns, thus the K band is of fundamental importance to measure the Sun-as-a-star radial velocity variations in the infrared domain. In the case of distant stars, this passband has been used to measure the RV and correct the variations induced by stellar activity, e.g., by means of the spectrograph CSHELL at NASA IRTF or NIRSPEC@Keck \cite{crockettetal2012}. Therefore, access to the K band is required to compare solar and stellar measurements.
Observation of the disc-integrated solar spectrum in the infrared K passband are crucial for an understanding of the effects of solar magnetic activity on the variations of the solar radial velocity. This happens because the K band samples layers of the solar atmosphere that are different from those that are sampled by the J and H passbands. Specifically, in the level sampled by the K passband, the brightness distribution, the velocity fields, and the magnetic field intensity are different than in those of the J and H passbands, allowing us a complete mapping of the photospheric convection and its perturbations in active regions (cf. \citenum{penn2014}). These results can be used to understand how magnetic fields affect the radial velocity measurements of the Sun as a star. Only in the Sun, we can compare disk-integrated measurements with spatially resolved maps of the active regions to test our models of the radial velocity perturbations to be used to correct activity effects in distant stars. The availability of data in the K passband is crucial to improve those models. 
Moreover, the Ti line at 2231 nm is an excellent probe of the relatively cooler plasma in sunspots and can be used to measure their effects on the disc-integrated flux, velocity fields, and mean magnetic field (e.g., \citenum{pennetal2003}).

   \begin{figure} [t]
   \begin{center}
%   \begin{tabular}{c} %% tabular useful for creating an array of images 
   \includegraphics[height=10 cm]{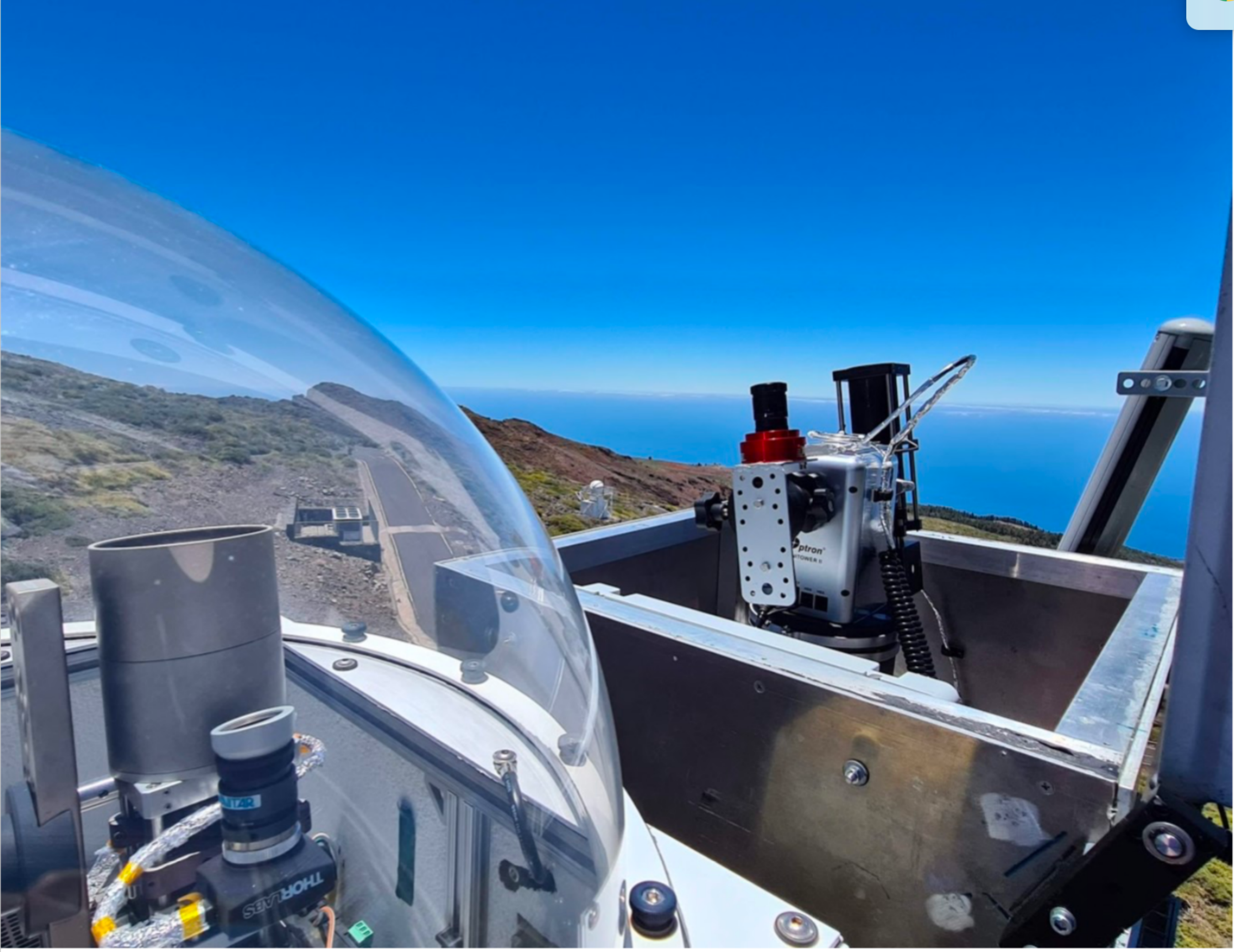}
%   \includegraphics{locnes_lcst.pdf}
%   \end{tabular}
   \end{center}
   \caption[example] 
%>>>> use \label inside caption to get Fig. number with \ref{}
   { \label{fig:tng} 
Mounted to West side of the plexiglass dome of LCST, the open dome of LOCNES shows the telescope, the guide camera and part of the mount}
   \end{figure} 

\section{LOCNES AT TNG}
\label{sec:locnes}
The whole LOCNES system is described in Figure\ \ref{fig:scheme} where, besides the telescope focusing the light into the integrating sphere, there are the mounts and the rest of equipment needed for the control of the system. 
The custom made dome is located on the south face of the TNG dome, close to the LCST dome (see Figure\ \ref{fig:tng}). The dome opens towards the North with a mechanical actuator and allows the telescope to be exposed for observations. The inside of the dome is equipped with an electronic box with power distribution, relays for opening/closing, communications to and from the control PC, surge protectors, environmental sensors and a webcam. 
The Mount is a commercial AZ-PRO from iOptron used in Alt/AZ mode. On the mount a lens with 1" aperture and 200mm focal length feeds the light of the Sun into the integrating sphere and constitutes the LOCNES optical tube assembly. The telescope is coupled in parallel with a 120MC guide camera from ZWO ASI.
From the integrating sphere 2 patches of optical fibers feed the scrambled light of the Sun to the TNG spectrographs. The fiber to Giano-B is a serie of two 20m long ZBLAN fibers with 200um diameter. Another 40m long LOW-OH fiber FG200LEA is pulled in order to be able to feed HARPS-N for tests and comparison in the output between LCST and LOCNES.
The control of the Mount, of the AG camera and of the Dome is made through an USB-over-fiber connection. A Black Box close to the control PC and one inside the dome are connected with a third 40m long optical fiber. The USB connection are used also for the webcam which allows to have a look inside the dome and for the arduino controlling the aperture of the dome and the environment (RH, Temp, vibrations).

\section{LOCNES EFFICIENCY}
\label{ssec:schedule}
To evaluate the apparent magnitude of the Sun once that its light has been collected by LOCNES, we can compare the efficiency of the solar telescope with that of TNG. This allows us to use the ETC of GIANO-B to evaluate the working exposure time and the value of the attenuation of the neutral filter, if any, that is necessary to use to avoid the saturation of the detector.
To do that we have to consider the output signal of LOCNES once it gathers the light from the Sun with its apparent magnitude $m_\odot$ and makes it equal to the output signal of TNG taken by a star with an equivalent magnitude $m_0$. The difference $m_\odot - m_0$ will be a function of the ratio of the efficiency of the two different paths of the two telescopes. 

\begin{figure}[h]
\begin{center}
%\sidecaption
% For example, with the graphicx style use
\includegraphics[scale=.40]{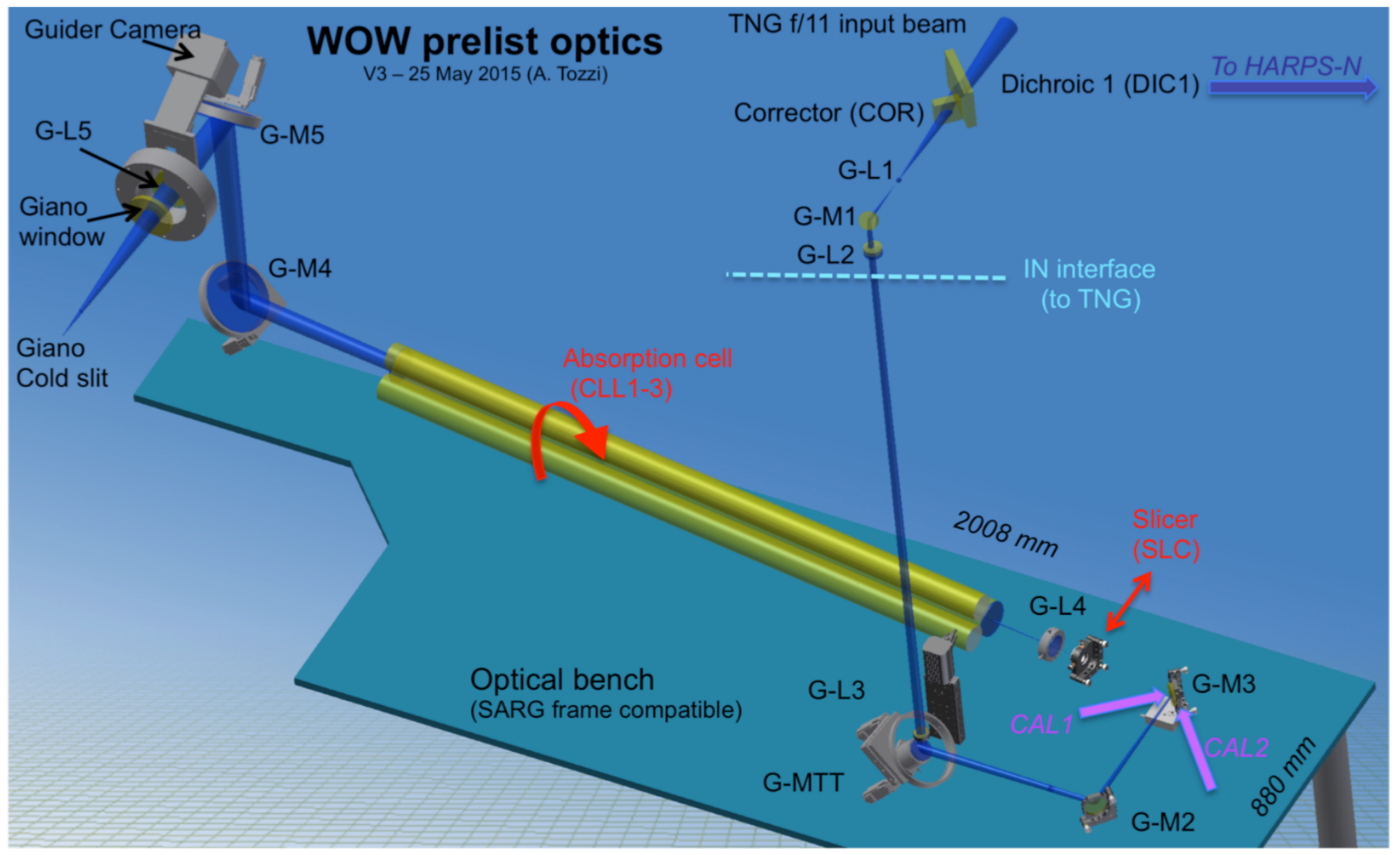}
%
% If no graphics program available, insert a blank space i.e. use
%\picplace{5cm}{2cm} % Give the correct figure height and width in cm
%
\caption{The GIANO-B Optical Train. In the sketch the optical path from the TNG is indicated. The LOCNES Fiber enters the Calibration Unit and illuminates the G-M2.}
\label{fig:gianob}       % Give a unique label
\end{center}
\end{figure}

The optical train of the TNG towards the high-resolution NIR spectrograph GIANO-B is constituted by the primary mirror (D=3.58 m), which acts as entrance pupil, and the secondary and tertiary mirrors, which inside their respective support barrels and baffles produce and obstruction of $\text{D}=1.2$\ m. The tertiary mirror folds the light beam toward the Nasmyth-B focus. At the focal plane, the light is intercepted by an optical relay called pick-up that folds the beam inside the GIANO-B pre-slit. Following this path, the light meets the following optical devices of the GIANO-B pre-slit (see Figure\ \ref{fig:gianob}): G-L1; G-M1; G-L2; G-MTT; G-M2.
LOCNES has a primary aperture defined by a lens with a diameter of $D_{lens}=0.0254$\ m. The lens insists on an Integrating Sphere (IS) with an internal diameter of $D_{IS}=0.0508$\ m that feed a double flight (20 m each) of Z-BLAN fiber. The fiber is thus 40 m long, with a connection in the middle. The core aperture is of 200 $\mu$m and with a Numerical Aperture of NA=0.2. The fiber attenuation is 0.25\ dB/m, while the conjunction loses about 1dB. Once that the fiber is inserted into the calibration unit, from G-M2 on the optical path follows the same route as the light coming from the TNG (see Figure\ \ref{fig:gianob}).
The signal collected by TNG is the following:
\begin{equation}
S_{TNG}=(1-\phi^2)A_{TNG}  \epsilon_1  \epsilon_2  \epsilon_{GIANO-B}  F_0 10^{-0.4 m_0}  t_{exp}
\label{eq:eftng}
\end{equation}

\noindent
where $(1-\phi^2)$ is the central obstruction of the telescope, $A_{TNG}$ is the entrance pupil area. The efficiency is split into three contributions that take into consideration: the optical train of TNG before the G-M2 ($\epsilon_1$), the remaining part of the GIANO-B pre-slit before the GIANO-B slit ($\epsilon_2$) and the efficiency of GIANO-B ($\epsilon_{GIANO-B}$). The $\epsilon_1$ is the product of the contribution of the three mirrors of the TNG (M$_1$; M$_2$, M$_3$) and the contribution of the pick-up and other optical elements: G-L1, G-M1, G-L2, G-L3, G-MTT, G-M2.

In the case of LOCNES instead, we have that the signal collected is the following:

\begin{equation}
S_{LOCNES}=A_{LOCNES}  \epsilon_{lens}  \epsilon_{is}  \epsilon_{fiber} \epsilon_{Ill}  \epsilon_2  \epsilon_{GIANO-B} F_0 10^{-0.4 m_\odot}  t_{exp}
\label{eq:eflocnes}
\end{equation}

\noindent
here $\epsilon_2$ and $\epsilon_{GIANO-B}$ are the same of those in Eq.\ \ref{eq:eftng}. Considering $S_{TNG}=S_{LOCNES}$, after a little of math, we have:

\begin{equation}
m_0 - m_\odot = -2.5 \log \left(\frac{A_{LOCNES} \epsilon_{lens} \epsilon_{is} \epsilon_{fiber} \epsilon_{Ill}}{(1-\phi^2)A_{TNG} \epsilon_1}\right)
\label{eq:ratio}
\end{equation}
The collecting area of the TNG is given by $(1-\phi^2)A_{TNG}$, where $\phi$ is the ration of the obstruction diameter (D$_{obs}=1.2$\ m) to the primary mirror diameter (D$_1=3.58$\ m), while $A_{TNG}=(\pi/4) D_1^2$.
To evaluate the resulting efficiency of the optical path through the not-common-route of the light from the TNG, we have to take into consideration the three reflections on the three telescope's mirrors and eventually the reflectivity and transmissivity of the other optical elements, G-L1, G-M1, G-L2, G-L3, and G-MTT. At the first approximation, we can consider a reflectivity of 0.90 for each mirror and a transmissivity of 0.90 for each lens. In the end, the total efficiency of the not-common-route optical path from TNG is $\epsilon_{NCP}=0.43$, while the collector area of the TNG is 8.935\ m$^2$.
As the LOCNES efficiency concern, things are a little bit more complicated, due to the presence of the integrating sphere coupled with the ZBLAN fibers. The input flux into the fiber is a function of the numerical aperture (NA) of the fiber, the area of the fiber core ($A_f$), and the reflectance losses to the air/fiber interface (R). The input flux is given by the following (Technical Note LabSphere\footnote{www.labsphere.com}):

\begin{equation}
\Phi_f= L_S A_f \pi (NA)^2 (1-R)\ \text{W}.
\label{eq:inpfib}
\end{equation}

\noindent
$L_S$ is the sphere surface radiance that is given by the following (Technical Note LabSphere):

\begin{equation}
L_s = \frac{\Phi_i}{\pi A_s} \frac{\rho}{1-\rho (1-f)} \text{W/m}^2\text{/sr}
\label{eq:sphere}
\end{equation}

\noindent
where: $\Phi_i$ is the input flux, $A_S$ is the inner surface of the integrating sphere, $f$ is the port fraction: $f=(A_i-A_e)/A_S$, $A_i$ is the input door surface, $A_e$ the output door surface, $\rho$ is the integrating sphere wall reflectivity.

\begin{figure}[h]
\begin{center}
%\sidecaption
% For example, with the graphicx style use
\includegraphics[scale=.20]{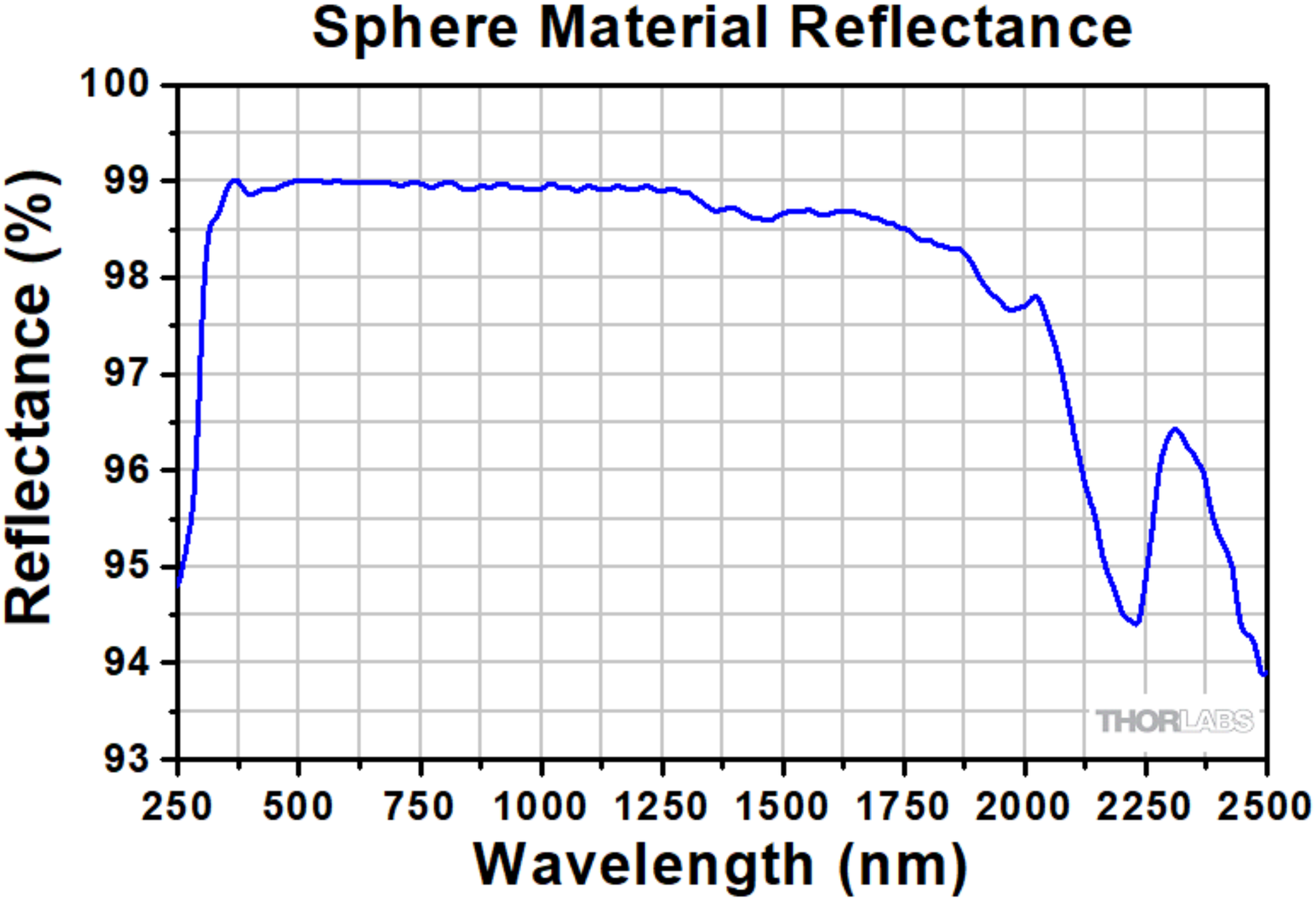}
\includegraphics[scale=.20]{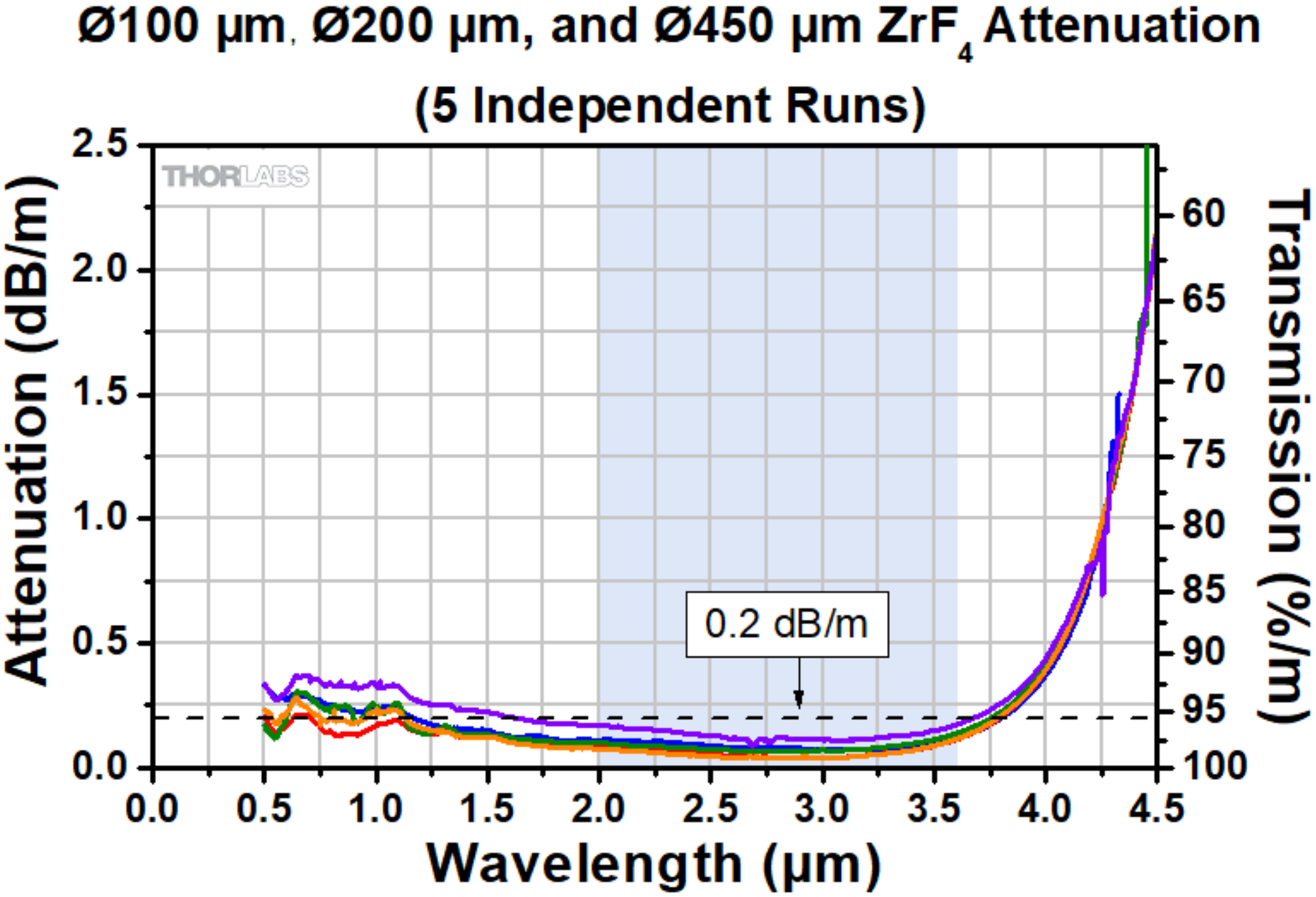}
\includegraphics[scale=.20]{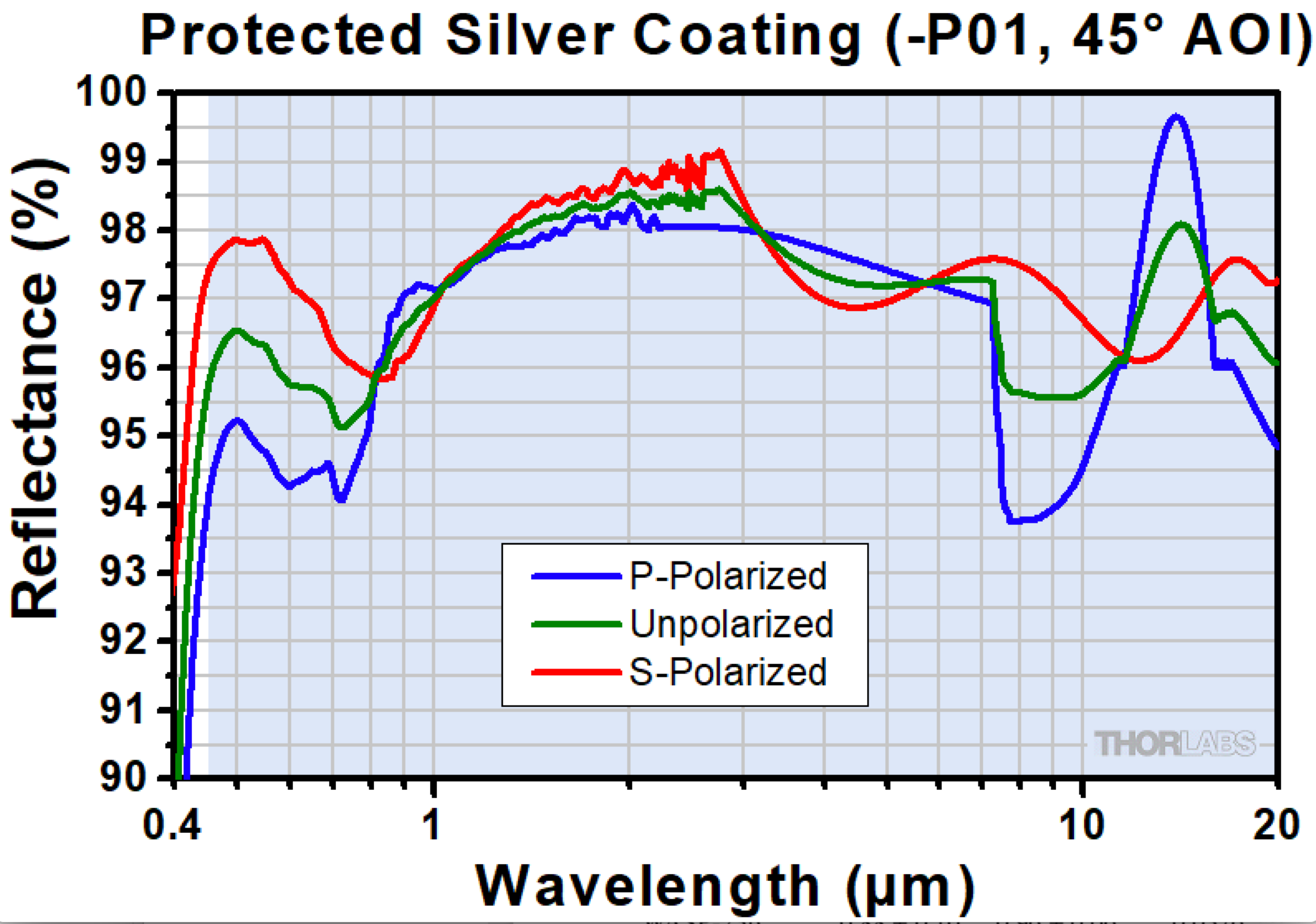}
%
% If no graphics program available, insert a blank space i.e. use
%\picplace{5cm}{2cm} % Give the correct figure height and width in cm
%
\caption{The factory data of the Integrating sphere, ZBLAN fibers, and off axis parabola. The latter is used in order to illuminate the GIANO-B slit.}
\label{fig:factdata}       % Give a unique label
\end{center}
\end{figure}

Considering the factory data (see Figure\ \ref{fig:factdata}) for all the optical components in the LOCNES branch before the injection of the light coming from LOCNES into the common-path part of the GIANO-B pre-slit, without entering into the details of the calculation, the Eq.\ \ref{eq:ratio} becomes:
\begin{equation}
m_0 - m_\odot =31.32\ \text{mag}
\label{eq:totallosslocnes}
\end{equation}

So, the equivalent magnitude of the Sun is: $m_0 = 31.32 + m_\odot $. We consider for GIANO-B the NIR band: J, H, K$_S$. Willmer (2018)(\citenum{willmer2018apjs}) gives the Sun apparent magnitude in several bands. We extracted the 2MASS band magnitudes. In Table\ \ref{tab:sunmag} these magnitudes are reported together with the LOCNES equivalent magnitude.

\begin{table}[h]
\caption{The apparent magnitude of the Sun \cite{willmer2018apjs} in the NIR bands of GIANO-B and the corresponding equivalent magnitude of the Sun through the LOCNES telescope.}
\begin{center}
\begin{tabular}{cccccccc}
\hline
2MASS     &   $\lambda_{mean} $   & FWHM   & m$_\odot$    & Equivalent &\multicolumn{3}{c}{SNR} \\
Band       &    ($\mu \text{m}$)        &                &          & mag&t$_{exp}=10$\ s & 30\ s& 60\ s \\
\hline
\hline
J              &    1.2411                    &0.2027    &  $-27.90$      & $3.42$ &85 & 150& 205 \\
H             &    1.6513                   & 0.2610     & $-28.50 $     & $2.82$ &115 &190 & 280$^\star$ \\
K$_S$             &     2.1650           & 0.2785      & $-28.30$    & $3.02$&90 & 150 &220 \\
\hline
\multicolumn{8}{l}{$^\star$Reached the non linear zone of the detector.}
\end{tabular}
\end{center}
\label{tab:sunmag}
\end{table}%

\section{Operation}
\label{ssec:operation}

\begin{figure}
\begin{center}
%\sidecaption
% For example, with the graphicx style use
\includegraphics[scale=.45]{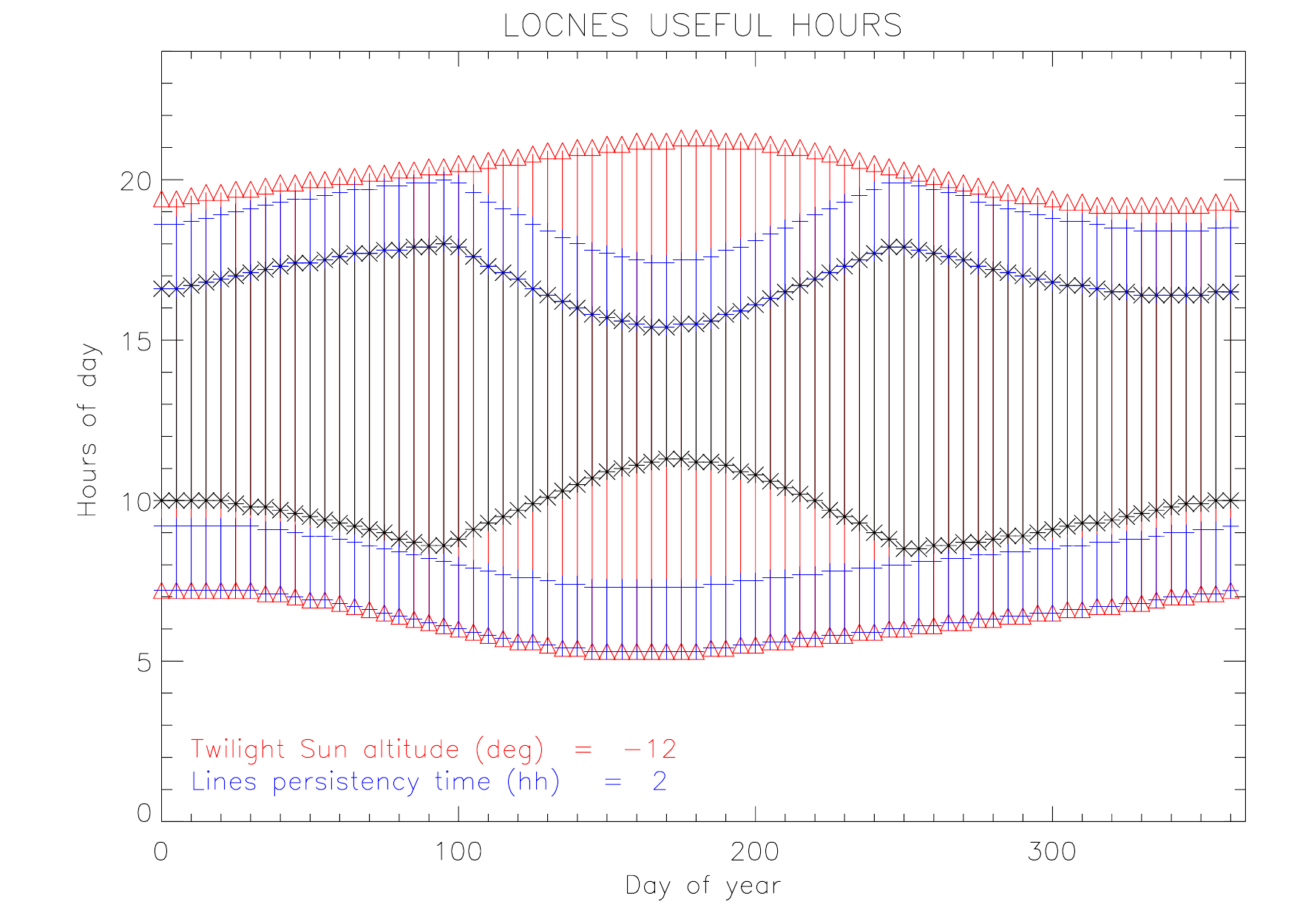}
\includegraphics[scale=.45]{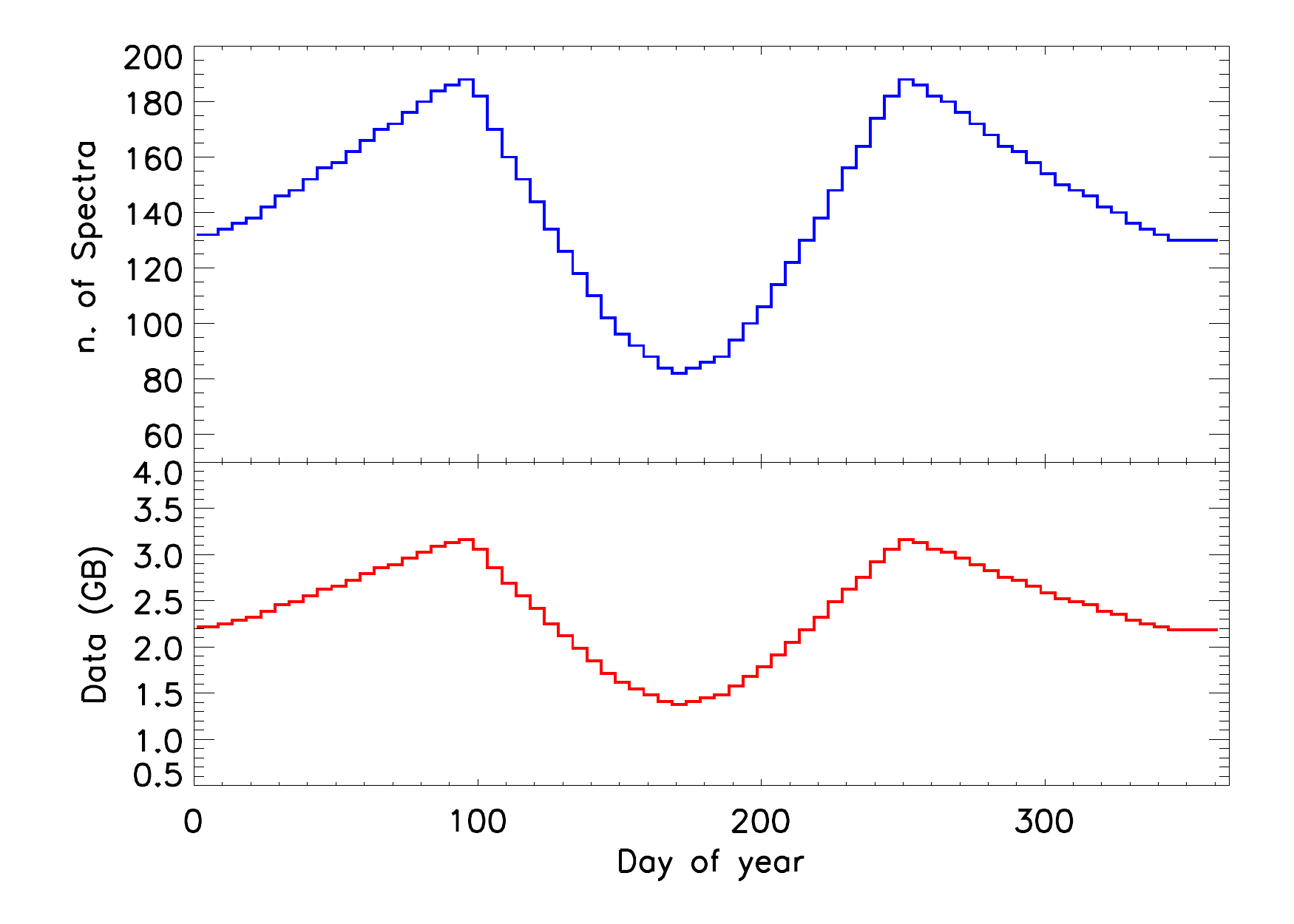}
%\includegraphics[scale=.40]{mettiilgraficotuo.pdf}
%
% If no graphics program available, insert a blank space i.e. use
%\picplace{5cm}{2cm} % Give the correct figure height and width in cm
%
\caption{The plot on the left shows the visibility of the Sun with LOCNES during the year (the black lines between asterisks). Starting from equinoxes the dome of the TNG limits the visibility to the East and to the West and this produces the hourglass shape in the plot. The red lines begins when the sun is at EL=$-12$ and this represents the end of the observing night, when calibrations are performed and thus persistence of the emission lines of lamps is possible: a delay of 2 hours after calibration is calculated in order to begin observations of the Sun when the detector is free from persistence. The right side figure shows the foreseen cumulative volume of data obtained during the year if LOCNES could always observe.}
\label{fig:oredisole}       % Give a unique label
\end{center}
\end{figure}

Once the opto-mechanical components will be fully installed and thoroughly tested, the system will be autonomous and will feed the light of the Sun every time that the weather will allow to open the dome. The observation of the Sun is limited in azimuth between East and West, due to the presence of the TNG dome, and in altitude above 20 degrees (to avoid the highest air-mass). These constraints imply that towards the summer time, even if the Sun is higher in the Sky, the hours that it can be observed are sometime less than in winter (see Figure\ \ref{fig:oredisole} left). 
LOCNES will be observing and feeding the light through the fiber independently to what GIANO-B is doing. It will let GIANO-B decide if it is ok to acquire spectra through some conditions like: \texttt{weather is OK}, \texttt{The sky is clear}, \texttt{the Sun is visible (EL,AZ)} , \texttt{the Sun is centered in the AG}.  
On the other side, at the end of a typical observing night of the TNG, GIANO-B will be prepared with a template to receive the light from LOCNES as if it were a calibration lamp, only that this time the lamp is our Sun. The spectra are continuously acquired with an exposure time based on the efficiency of the whole system, and depending on the time of the year, the maximum number of collectible spectra is foreseen as in Figure\ \ref{fig:oredisole} right.

\section{Sun's Radial Velocity measurements in the NIR Range}
\label{ssec:rv}
\begin{figure} [ht]
   \begin{center}
   \begin{tabular}{c} %% tabular useful for creating an array of images 
   \includegraphics[height=5cm]{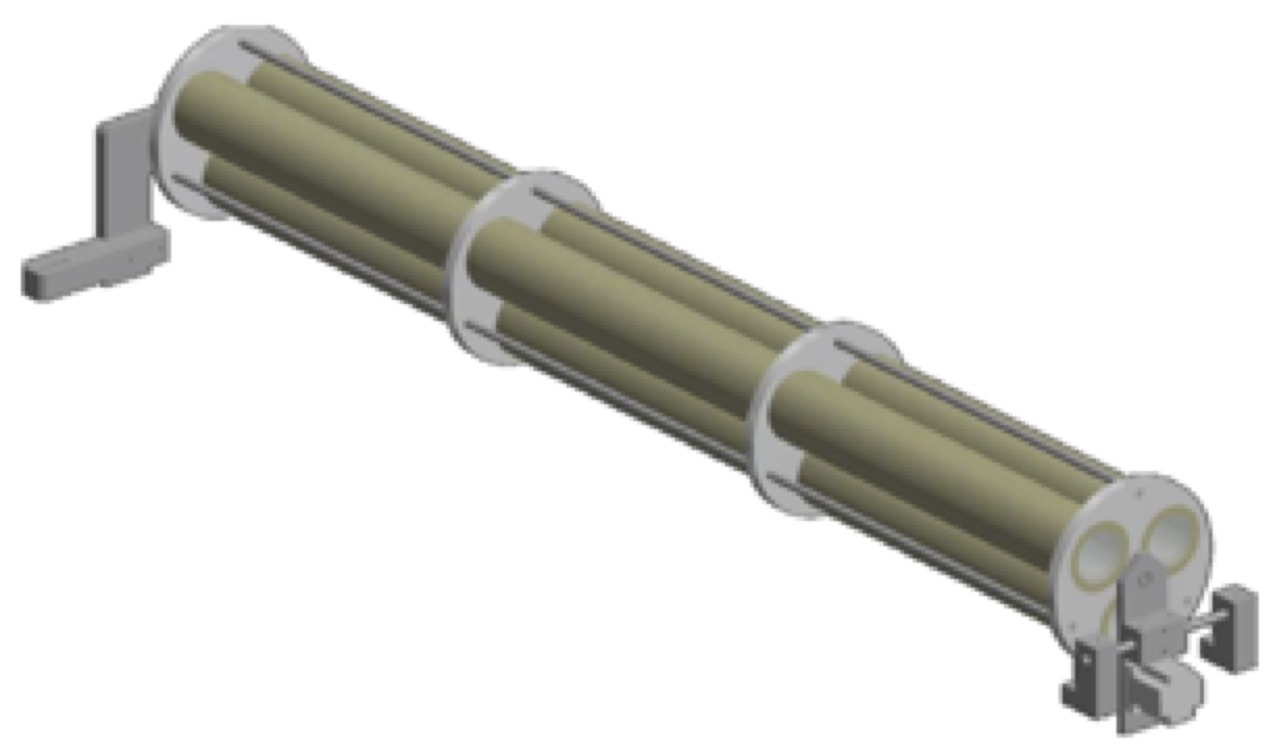}
   \end{tabular}
   \end{center}
   \caption[example] 
%>>>> use \label inside caption to get Fig. number with \ref{}
   { \label{fig:celle} 
Mechanical design for the GIANO cell revolver. The revolver mechanism mounts two cells (and an open slot), and rotates into the optical beam. The revolver is located in the preslit area of GIANO-B \cite{tozzietal2016} in a dedicated structure. }
   \end{figure}

GIANO-B will be equipped with NIR absorbing cells in order to have an inertial inner reference for high precision radial velocity measurements. The cells will be mounted on a revolver inside the preslit optics (Figure \ref{fig:celle}) to be able to put the absorbing cell into the optical path towards the GIANO--B slit.

In this way, a spectrum of CH$_4$-C$_2$H$_2$-NH$_3$, the gas mixture present into the absorbing cell \cite{seemanetal2018}, is over imposed to the stellar spectrum. The value of the radial velocity of the star is so evaluated by fit of the composite spectrum by a synthetic reconstruction of it using the cell spectrum, the instrumental profile of GIANO-B and a high signal to noise spectrum of the star alone. The absorbing cell technique is described in Butler et al (\citenum{butleretal1996}).

In this moment, the cells are under construction and we developed an alternative procedure to get the value of RVs by the telluric lines.
The method to extract radial velocities (RV) follows the approach described in Carleo et al (\citenum{carleoetal2016}): the telluric lines are used as wavelength reference and the cross-correlation function (CCF) method is used to determine the stellar RV. 

\section{First Sun-as-a-star Spectra}
\label{sec:sas}
Once that LOCNES has been fastened on the exterior stairs of the TNG Dome, the optical fiber patches were connected to the integrating sphere of LOCNES. The ZBLAN, VIS and COM fibers have been extended along the TNG structure down to the Nasmyth room towards the GIANO-B pre-slit. This operation has a level of risk for the ZBLAN fibers because they are very brittle fibers; for the same reason they are difficult to produce and we had to choose the $200\mu$m size instead of a $400\mu$m. 

\begin{figure} [ht]
   \begin{center}
   \begin{tabular}{c} %% tabular useful for creating an array of images 
   \includegraphics[height=8cm]{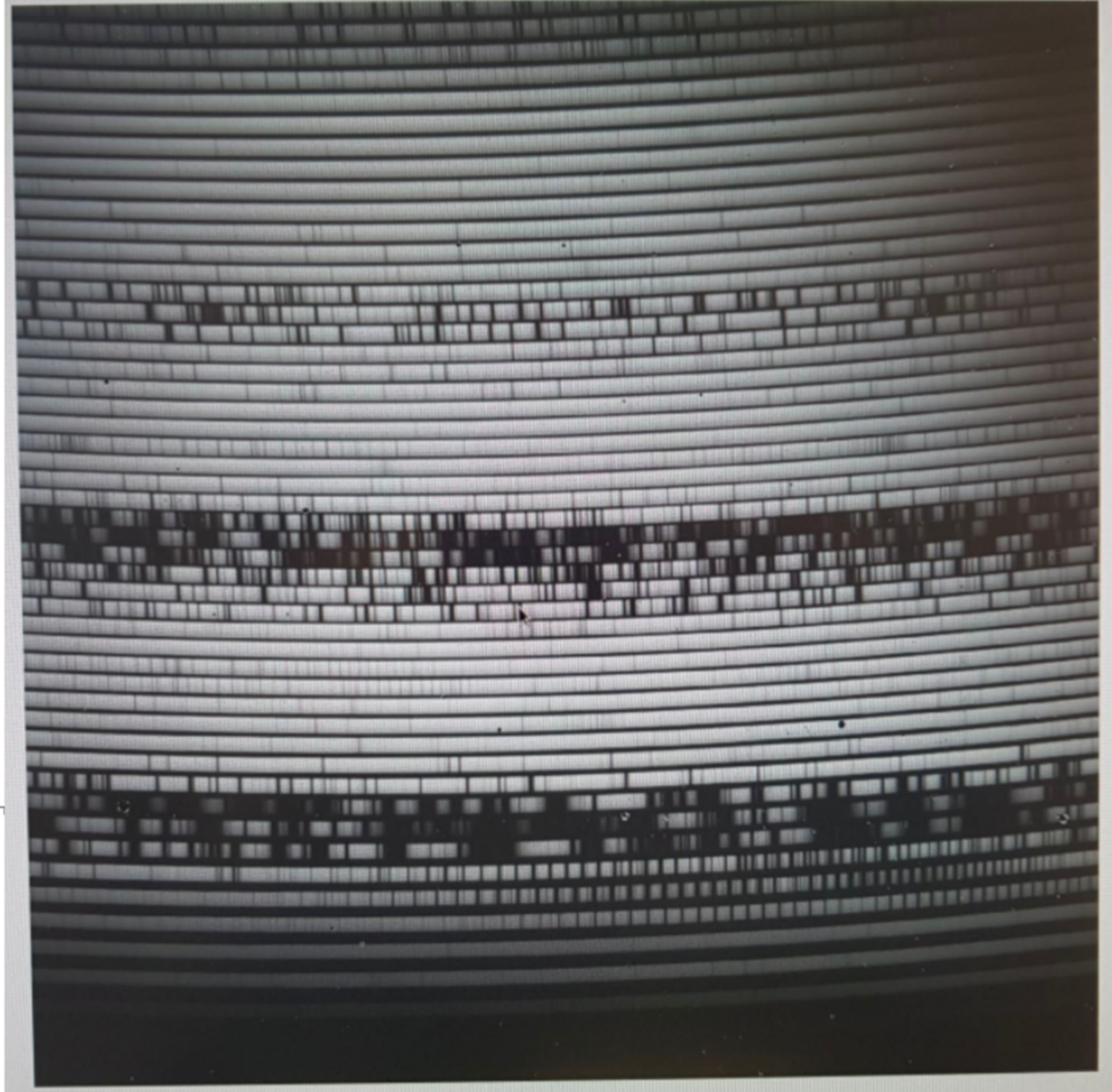}
   \includegraphics[height=8cm]{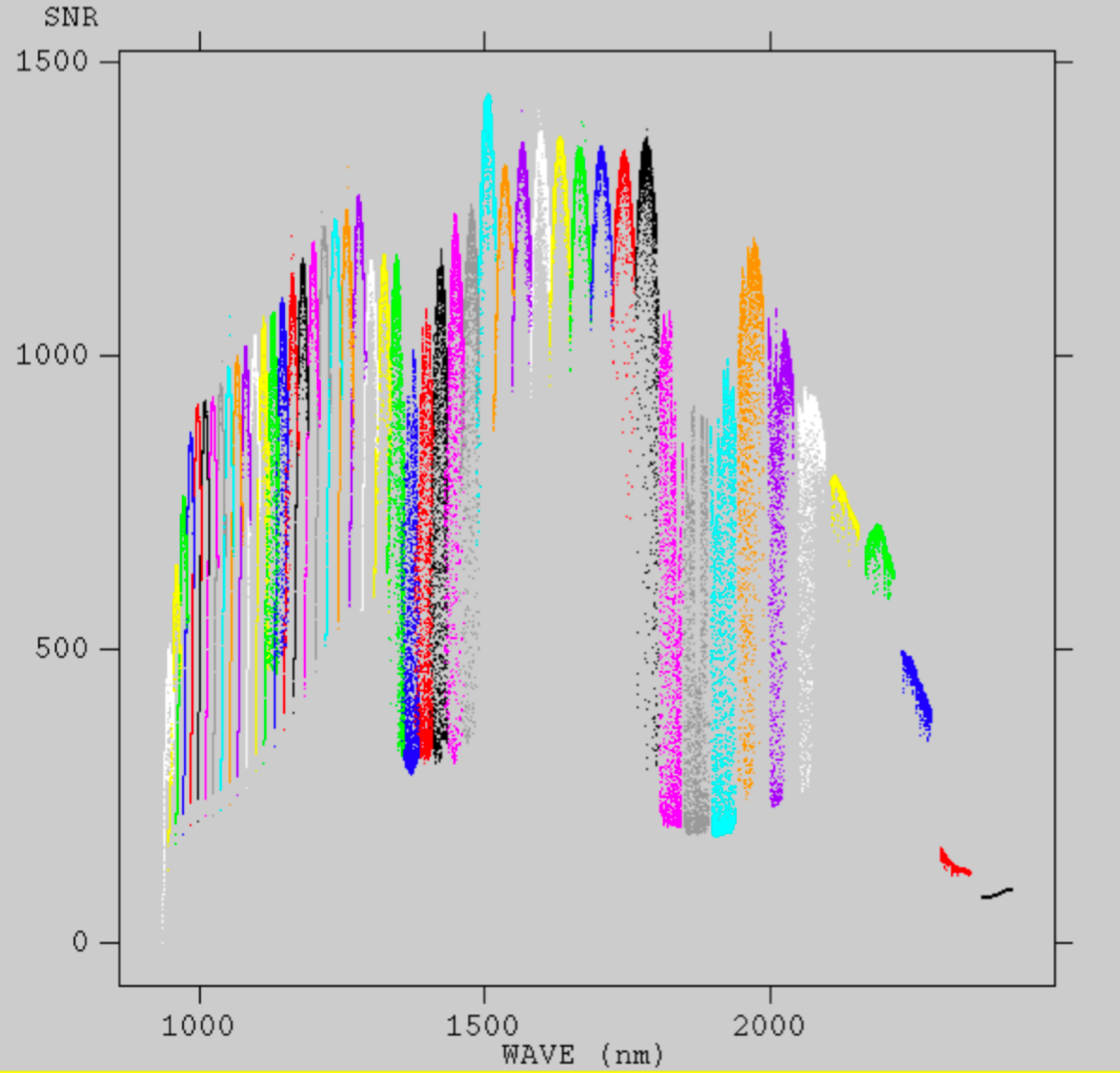}
   \end{tabular}
   \end{center}
   \caption[example] 
%>>>> use \label inside caption to get Fig. number with \ref{}
   { \label{fig:spectra} \textbf{Left}: GIANO-B spectral format;  \textbf{Right}: 1-D wavelength calibrated Solar spectrum. The value of the SNR is clearly shown. The drop above $2.2\mu$m is clearly visible in both images. To exploit in full the capabilities of GIANO-B we had to go with the ZBLAN fiber.
}
   \end{figure}

Eventually, before choosing the $\sim 10$ times more expensive ZBLAN fibers we tried to go with the Low-OH and see how bearable was the cut on the wavelengths in the NIR above $2.2\mu$m. We set up LOCNES with a Low-OH fiber and we were able to feed the pre-slit of GIANO-B and thus obtain the first set of NIR spectra of the Sun with LOCNES  (see Figure\ \ref{fig:spectra}). 
The drop due to the characteristics of the Low-OH fiber is clearly visible in the spectra and to achieve the scientific drivers of LOCNES the visibility of the lines in the K band is mandatory. We keep the Low-OH fiber for possible comparisons between LOCNES and LCST but we finally bought and installed the ZBLAN fiber to feed GIANO-B.

We use these spectra to test also the level of pollution by telluric lines during daily observations. 

Observations from the ground are plagued by contamination of Earth's atmosphere that introduces the so-called  telluric lines into the spectrum. In particular, the NIR domain is heavily affected from water vapour and O$_{2}$ lines, which dominate the signal.
The standard approach, consisting in the observation of a standard star at the same airmass and atmospheric condition (i.e., temperature, pressure and humidity), is obviously not applicable in our case. For this reason, we proceeded by calculating synthetic telluric spectra, which have been computed using TelFit \cite{gulliksonetal2014aj}. This object-oriented Python code is extremely flexible and the user can decide which atmospheric parameters to fit (in our case pressure and humidity since airmass is close to 1 in almost all our observations), along with the resolution and the wavelength solution (i.e., the algorithm allows to re-adjust possible shifts between model and observations).

  \begin{figure} [ht]
   \begin{center}
   \begin{tabular}{c} %% tabular useful for creating an array of images 
   \includegraphics[height=6cm]{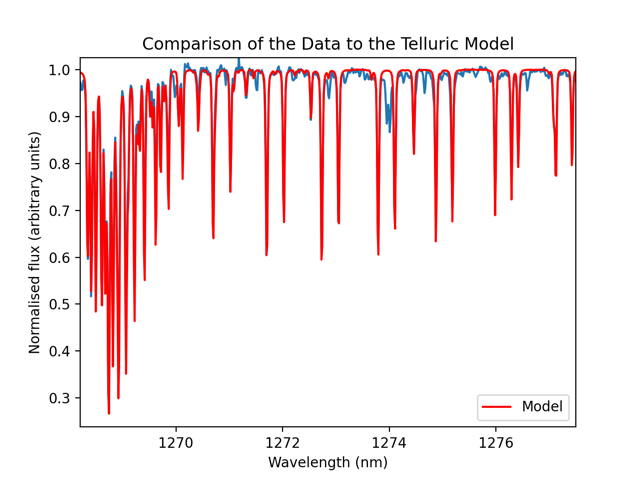}
   \includegraphics[height=6cm]{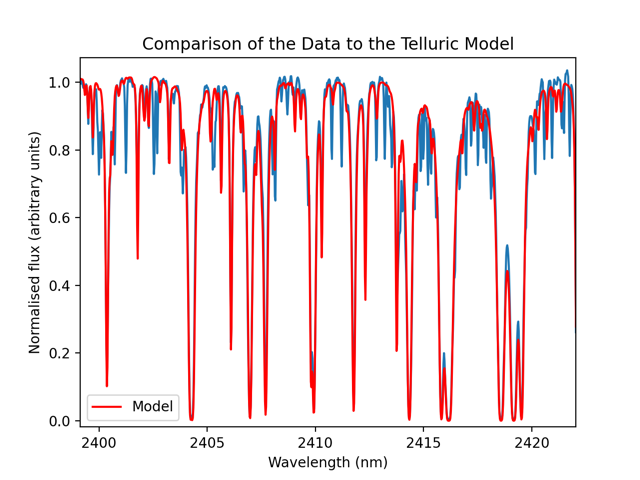}
   \end{tabular}
   \end{center}
   \caption[example] 
%>>>> use \label inside caption to get Fig. number with \ref{}
   { \label{fig:tell} \textbf{Left}: Observed Solar spectra (blue) and model of telluric spectra (red) in the J band. \textbf{Right}: The same but for the K band.
 }
   \end{figure} 
An example of the results we obtained is shown in Figures\ \ref{fig:tell}.
In the next-coming months, we plan to explore the use of the ESO standard software MOLECFIT \cite{kauschetal2015aa, smetteetal2015aa} and to implement the automatic procedure into the GOFIO reduction pipeline \cite{raineretal2018spie}.

\section{CONCLUSION}
\label{ssec:schedule}
We outlined the scientific aims and the description of the LOCNES telescope project. The wide wavelength range of LOCNES will help to connect the stellar (solar) activity causes and the different zones and depths of the chromosphere to the resulting RVs allowing the development of more tools for an optimal activity correction. Furthermore, the TNG acquires a simple, low-cost instrument that makes it the unique astronomical structure in the world with the possibility to gather high-resolution spectra of the Sun-as-a-star from the visible up to the K band. Furthermore, we gave a snapshot of the status of the project, its possibility, and efficiency. We also provided an evaluation of the mole of data and of the number of spectra that LOCNES can produce in one year.

\acknowledgments % equivalent to \section*{ACKNOWLEDGMENTS}       
 
 Authors would like to thank the support by INAF through Progetti Premiali funding scheme of the Italian Ministry of Education, University, and Research.

% References
\bibliography{report} % bibliography data in report.bib

\begin{thebibliography}{10}

\bibitem{dumusqueetal2015apjl}
{Dumusque}, X., {Glenday}, A., {Phillips}, D.~F., {Buchschacher}, N., {Collier
  Cameron}, A., {Cecconi}, M., {Charbonneau}, D., {Cosentino}, R., {Ghedina},
  A., {Latham}, D.~W., {Li}, C.-H., {Lodi}, M., {Lovis}, C., {Molinari}, E.,
  {Pepe}, F., {Udry}, S., {Sasselov}, D., {Szentgyorgyi}, A., and {Walsworth},
  R., ``Harps-n observes the sun as a star,'' {\em ApJL}~{\bf 814},  L21
  (2015).

\bibitem{marchwinskietal2015}
{Marchwinski}, R.~C., {Mahadevan}, S., {Robertson}, P., {Ramsey}, L., and
  {Harder}, J., ``{Toward Understanding Stellar Radial Velocity Jitter as a
  Function of Wavelength: The Sun as a Proxy},'' {\em ApJ}~{\bf 798},  63
  (2015).

\bibitem{haywoodetal2016}
{Haywood}, R.~D., {Collier Cameron}, A., {Unruh}, Y.~C., {Lovis}, C., {Lanza},
  A.~F., {Llama}, J., {Deleuil}, M., {Fares}, R., {Gillon}, M., {Moutou}, C.,
  {Pepe}, F., {Pollacco}, D., {Queloz}, D., and {S{\'e}gransan}, D., ``{The Sun
  as a planet-host star: proxies from SDO images for HARPS radial-velocity
  variations},'' {\em MNRAS}~{\bf 457},  3637--3651 (2016).

\bibitem{phillipsetal2016}
{Phillips}, D.~F., {Glenday}, A.~G., {Dumusque}, X., {Buchschacher}, N.,
  {Cameron}, A.~C., {Cecconi}, M., {Charbonneau}, D., {Cosentino}, R.,
  {Ghedina}, A., {Haywood}, R., {Latham}, D.~W., {Li}, C.-H., {Lodi}, M.,
  {Lovis}, C., {Molinari}, E., {Pepe}, F., {Sasselov}, D., {Szentgyorgyi}, A.,
  {Udry}, S., and {Walsworth}, R.~L., ``An astro-comb calibrated solar
  telescope to search for the radial velocity signature of venus,'' in [{\em
  Advances in Optical and Mechanical Technologies for Telescopes and
  Instrumentation II}{\nolinebreak\hspace{0.1em}]},  {\em Proc. SPIE} {\bf
  9912},  99126Z (2016).

\bibitem{dumusqueetal2020arxiv}
{Dumusque}, X., {Cretignier}, M., {Sosnowska}, D., {Buchschacher}, N., {Lovis},
  C., {Phillips}, D.~F., {Pepe}, F., {Alesina}, F., {Buchhave}, L.~A.,
  {Burnier}, J., {Cecconi}, M., {Cegla}, H.~M., {Cloutier}, R., {Collier
  Cameron}, A., {Cosentino}, R., {Ghedina}, A., {Gonzalez}, M., {Haywood},
  R.~D., {Latham}, D.~W., {Lodi}, M., {Lopez-Morales}, M., {Maldonado}, J.,
  {Malavolta}, L., {Micela}, G., {Molinari}, E., {Mortier}, A., {Perez
  Ventura}, H., {Pinamonti}, M., {Poretti}, E., {Rice}, K., {Riverol}, L.,
  {Riverol}, C., {San Juan}, J., {Segransan}, D., {Sozzetti}, A., {Thompson},
  S.~J., {Udry}, S., and {Wilson}, T.~G., ``{Three Years of HARPS-N
  High-Resolution Spectroscopy and Precise Radial Velocity Data for the Sun},''
  {\em arXiv e-prints} ,  arXiv:2009.01945 (Sept. 2020).

\bibitem{colliercameron2019mnras}
{Collier Cameron}, A., {Mortier}, A., {Phillips}, D., {Dumusque}, X.,
  {Haywood}, R.~D., {Langellier}, N., {Watson}, C.~A., {Cegla}, H.~M.,
  {Costes}, J., {Charbonneau}, D., {Coffinet}, A., {Latham}, D.~W.,
  {Lopez-Morales}, M., {Malavolta}, L., {Maldonado}, J., {Micela}, G.,
  {Milbourne}, T., {Molinari}, E., {Saar}, S.~H., {Thompson}, S.,
  {Buchschacher}, N., {Cecconi}, M., {Cosentino}, R., {Ghedina}, A., {Glenday},
  A., {Gonzalez}, M., {Li}, C.~H., {Lodi}, M., {Lovis}, C., {Pepe}, F.,
  {Poretti}, E., {Rice}, K., {Sasselov}, D., {Sozzetti}, A., {Szentgyorgyi},
  A., {Udry}, S., and {Walsworth}, R., ``{Three years of Sun-as-a-star
  radial-velocity observations on the approach to solar minimum},'' {\em
  \mnras}~{\bf 487},  1082--1100 (July 2019).

\bibitem{milbourneetal2019apj}
{Milbourne}, T.~W., {Haywood}, R.~D., {Phillips}, D.~F., {Saar}, S.~H.,
  {Cegla}, H.~M., {Cameron}, A.~C., {Costes}, J., {Dumusque}, X., {Langellier},
  N., {Latham}, D.~W., {Maldonado}, J., {Malavolta}, L., {Mortier}, A.,
  {Palumbo}, M.~L., I., {Thompson}, S., {Watson}, C.~A., {Bouchy}, F.,
  {Buchschacher}, N., {Cecconi}, M., {Charbonneau}, D., {Cosentino}, R.,
  {Ghedina}, A., {Glenday}, A.~G., {Gonzalez}, M., {Li}, C.~H., {Lodi}, M.,
  {L{\'o}pez-Morales}, M., {Lovis}, C., {Mayor}, M., {Micela}, G., {Molinari},
  E., {Pepe}, F., {Piotto}, G., {Rice}, K., {Sasselov}, D., {S{\'e}gransan},
  D., {Sozzetti}, A., {Szentgyorgyi}, A., {Udry}, S., and {Walsworth}, R.~L.,
  ``{HARPS-N Solar RVs Are Dominated by Large, Bright Magnetic Regions},'' {\em
  \apj}~{\bf 874},  107 (Mar. 2019).

\bibitem{maldonadoetal2019aa}
{Maldonado}, J., {Phillips}, D.~F., {Dumusque}, X., {Collier Cameron}, A.,
  {Haywood}, R.~D., {Lanza}, A.~F., {Micela}, G., {Mortier}, A., {Saar}, S.~H.,
  {Sozzetti}, A., {Rice}, K., {Milbourne}, T., {Cecconi}, M., {Cegla}, H.~M.,
  {Cosentino}, R., {Costes}, J., {Ghedina}, A., {Gonzalez}, M., {Guerra}, J.,
  {Hern{\'a}ndez}, N., {Li}, C.~H., {Lodi}, M., {Malavolta}, L., {Molinari},
  E., {Pepe}, F., {Piotto}, G., {Poretti}, E., {Sasselov}, D., {San Juan}, J.,
  {Thompson}, S., {Udry}, S., and {Watson}, C., ``{Temporal evolution and
  correlations of optical activity indicators measured in Sun-as-a-star
  observations},'' {\em \aap}~{\bf 627},  A118 (July 2019).

\bibitem{miklosetal2020apj}
{Miklos}, M., {Milbourne}, T.~W., {Haywood}, R.~D., {Phillips}, D.~F., {Saar},
  S.~H., {Meunier}, N., {Cegla}, H.~M., {Dumusque}, X., {Langellier}, N.,
  {Maldonado}, J., {Malavolta}, L., {Mortier}, A., {Thompson}, S., {Watson},
  C.~A., {Cecconi}, M., {Cosentino}, R., {Ghedina}, A., {Li}, C.~H.,
  {L{\'o}pez-Morales}, M., {Molinari}, E., {Poretti}, E., {Sasselov}, D.,
  {Sozzetti}, A., and {Walsworth}, R.~L., ``{Testing the Spectroscopic
  Extraction of Suppression of Convective Blueshift},'' {\em \apj}~{\bf 888},
  117 (Jan. 2020).

\bibitem{langellieretal2020arxiv}
{Langellier}, N., {Milbourne}, T.~W., {Phillips}, D.~F., {Haywood}, R.~D.,
  {Saar}, S.~H., {Mortier}, A., {Malavolta}, L., {Thompson}, S., {Collier
  Cameron}, A., {Dumusque}, X., {Cegla}, H.~M., {Latham}, D.~W., {Maldonado},
  J., {Watson}, C.~A., {Cecconi}, M., {Charbonneau}, D., {Cosentino}, R.,
  {Ghedina}, A., {Gonzalez}, M., {Li}, C.-H., {Lodi}, M., {L{\'o}pez-Morales},
  M., {Micela}, G., {Molinari}, E., {Pepe}, F., {Poretti}, E., {Rice}, K.,
  {Sasselov}, D., {Sozzetti}, A., {Udry}, S., and {Walsworth}, R.~L.,
  ``{Detection Limits of Low-mass, Long-period Exoplanets Using Gaussian
  Processes Applied to HARPS-N Solar RVs},'' {\em arXiv e-prints} ,
  arXiv:2008.05970 (Aug. 2020).

\bibitem{olivaetal2012}
{Oliva}, E., {Origlia}, L., {Maiolino}, R., {Baffa}, C., {Biliotti}, V.,
  {Bruno}, P., {Falcini}, G., {Gavriousev}, V., {Ghinassi}, F., {Giani}, E.,
  {Gonzalez}, M., {Leone}, F., {Lodi}, M., {Massi}, F., {Mochi}, I.,
  {Montegriffo}, P., {Pedani}, M., {Rossetti}, E., {Scuderi}, S., {Sozzi}, M.,
  and {Tozzi}, A., ``The giano spectrometer: towards its first light at the
  tng,'' in [{\em Ground-based and Airborne Instrumentation for Astronomy
  IV}{\nolinebreak\hspace{0.1em}]},  {\em Proc. SPIE} {\bf 8446},  84463T
  (2012).

\bibitem{claudietal2017}
{Claudi}, R., {Benatti}, S., {Carleo}, I., {Ghedina}, A., {Guerra}, J.,
  {Micela}, G., {Molinari}, E., {Oliva}, E., {Rainer}, M., {Tozzi}, A.,
  {Baffa}, C., {Baruffolo}, A., {Buchschacher}, N., {Cecconi}, M., {Cosentino},
  R., {Fantinel}, D., {Fini}, L., {Ghinassi}, F., {Giani}, E., {Gonzalez}, E.,
  {Gonzalez}, M., {Gratton}, R., {Harutyunyan}, A., {Hernandez}, N., {Lodi},
  M., {Malavolta}, L., {Maldonado}, J., {Origlia}, L., {Sanna}, N., {Sanjuan},
  J., {Scuderi}, S., {Seemann}, U., {Sozzetti}, A., {Perez Ventura}, H.,
  {Hernandez Diaz}, M., {Galli}, A., {Gonzalez}, C., {Riverol}, L., and
  {Riverol}, C., ``Giarps@tng: Giano-b and harps-n together for a wider
  wavelength range spectroscopy,'' {\em European Physical Journal Plus}~{\bf
  132},  364 (2017).

\bibitem{claudietal2018spie}
{Claudi}, R., {Benatti}, S., {Carleo}, I., {Ghedina}, A., {Guerra}, J.,
  {Ghinassi}, F., {Harutyunyan}, A., {Micela}, G., {Molinari}, E., {Oliva}, E.,
  {Rainer}, M., {Tozzi}, A., {Baffa}, C., {Baruffolo}, A., {Biliotti}, V.,
  {Buchschacher}, N., {Cecconi}, M., {Cosentino}, R., {Falcini}, G.,
  {Fantinel}, D., {Fini}, L., {Giani}, E., {Gonzalez-Alvarez}, E., {Gonzalez},
  M., {Gonzalez}, C., {Gratton}, R., {Hernandez}, N., {Iuzzolino}, M., {Lodi},
  M., {Malavolta}, L., {Maldonado}, J., {Origlia}, L., {Puglisi}, A., {Sanna},
  N., {San Juan G{\'o}mez}, J., {Scuderi}, S., {Seemann}, U., {Sozzetti}, A.,
  {Sozzi}, M., {Perez Ventura}, H., {Hernandez Diaz}, M., {Galli}, A.,
  {Riverol}, L., and {Riverol}, C., ``{GIARPS: commissioning and first
  scientific results},'' in [{\em Ground-based and Airborne Instrumentation for
  Astronomy VII}{\nolinebreak\hspace{0.1em}]},  {Evans}, C.~J., {Simard}, L.,
  and {Takami}, H., eds., {\em Society of Photo-Optical Instrumentation
  Engineers (SPIE) Conference Series} {\bf 10702},  107020Z (July 2018).

\bibitem{boisseetal2009}
{Boisse}, I., {Moutou}, C., {Vidal-Madjar}, A., {Bouchy}, F., {Pont}, F.,
  {H{\'e}brard}, G., {Bonfils}, X., {Croll}, B., {Delfosse}, X., {Desort}, M.,
  {Forveille}, T., {Lagrange}, A.-M., {Loeillet}, B., {Lovis}, C., {Matthews},
  J.~M., {Mayor}, M., {Pepe}, F., {Perrier}, C., {Queloz}, D., {Rowe}, J.~F.,
  {Santos}, N.~C., {S{\'e}gransan}, D., and {Udry}, S., ``{Stellar activity of
  planetary host star HD 189 733},'' {\em A\&A}~{\bf 495},  959--966 (2009).

\bibitem{aigrainetal2012}
{Aigrain}, S., {Pont}, F., and {Zucker}, S., ``{A simple method to estimate
  radial velocity variations due to stellar activity using photometry},'' {\em
  MNRAS}~{\bf 419},  3147--3158 (2012).

\bibitem{vogt1987}
{Vogt}, S.~S., ``{Doppler images of rotating stars from high precision
  spectroscopy},'' in [{\em Liege International Astrophysical
  Colloquia}{\nolinebreak\hspace{0.1em}]},  {Swings}, J.-P., {Collin}, J., and
  {Wampler}, E.~J., eds., {\em Liege International Astrophysical Colloquia}
  {\bf 27},  317--325 (1987).

\bibitem{quelozetal2001}
{Queloz}, D., {Henry}, G.~W., {Sivan}, J.~P., {Baliunas}, S.~L., {Beuzit},
  J.~L., {Donahue}, R.~A., {Mayor}, M., {Naef}, D., {Perrier}, C., and {Udry},
  S., ``{No planet for HD 166435},'' {\em A\& A}~{\bf 379},  279--287 (2001).

\bibitem{figueiraetal2013}
{Figueira}, P., {Santos}, N.~C., {Pepe}, F., {Lovis}, C., and {Nardetto}, N.,
  ``{Line-profile variations in radial-velocity measurements. Two alternative
  indicators for planetary searches},'' {\em A \& A,}~{\bf 557},  A93 (2013).

\bibitem{dumusqueetal2014}
{Dumusque}, X., {Boisse}, I., and {Santos}, N.~C., ``{SOAP 2.0: A Tool to
  Estimate the Photometric and Radial Velocity Variations Induced by Stellar
  Spots and Plages},'' {\em ApJ,}~{\bf 796},  132 (2014).

\bibitem{noyesetal1984}
{Noyes}, R.~W., {Weiss}, N.~O., and {Vaughan}, A.~H., ``{The relation between
  stellar rotation rate and activity cycle periods},'' {\em ApJ,}~{\bf 287},
  769--773 (1984).

\bibitem{pesnelletal2012}
{Pesnell}, W.~D., {Thompson}, B.~J., and {Chamberlin}, P.~C., ``{The Solar
  Dynamics Observatory (SDO)},'' {\em Sol. Phys.,}~{\bf 275},  3--15 (2012).

\bibitem{demingandplymate1994}
{Deming}, D. and {Plymate}, C., ``{On the apparent velocity of integrated
  sunlight. 2: 1983-1992 and comparisons with magnetograms},'' {\em ApJ,}~{\bf
  426},  382--386 (1994).

\bibitem{crockettetal2012}
{Crockett}, C.~J., {Mahmud}, N.~I., {Prato}, L., {Johns-Krull}, C.~M., {Jaffe},
  D.~T., {Hartigan}, P.~M., and {Beichman}, C.~A., ``{A Search for Giant Planet
  Companions to T Tauri Stars},'' {\em ApJ,}~{\bf 761},  164 (2012).

\bibitem{penn2014}
{Penn}, M.~J., ``{Infrared Solar Physics},'' {\em Living Reviews in Solar
  Physics,}~{\bf 11},  2 (2014).

\bibitem{pennetal2003}
{Penn}, M.~J., {Cao}, W.~D., {Walton}, S.~R., {Chapman}, G.~A., and
  {Livingston}, W., ``{Imaging Spectropolarimetry of Ti I 2231 nm in a
  Sunspot},'' {\em Sol. Phys.,}~{\bf 215},  87--97 (2003).

\bibitem{willmer2018apjs}
{Willmer}, C. N.~A., ``{The Absolute Magnitude of the Sun in Several
  Filters},'' {\em \apjs}~{\bf 236},  47 (June 2018).

\bibitem{seemanetal2018}
{Seemann}, U.~e., ``{The TNG/GIARPS gas-absorption cell for near-infrared
  precision radial velocities},'' in [{\em Ground-based and Airborne
  Instrumentation for Astronomy 2018}{\nolinebreak\hspace{0.1em}]},  {\em Proc.
  SPIE} {\bf 10706-235,} (2018).

\bibitem{butleretal1996}
{Butler}, R.~P., {Marcy}, G.~W., {Williams}, E., {McCarthy}, C., {Dosanjh}, P.,
  and {Vogt}, S.~S., ``{Attaining Doppler Precision of 3 M s-1},'' {\em
  PASP}~{\bf 108},  500 (1996).

\bibitem{tozzietal2016}
{Tozzi}, A., {Oliva}, E., {Iuzzolino}, M., {Fini}, L., {Puglisi}, A., {Sozzi},
  M., {Falcini}, G., {Carbonaro}, L., {Ghedina}, A., {Mercatelli}, L.,
  {Seemann}, U., and {Claudi}, R., ``{GIANO and HARPS-N together: towards an
  Earth-mass detection instrument},'' in [{\em Ground-based and Airborne
  Instrumentation for Astronomy VI}{\nolinebreak\hspace{0.1em}]},  {\em Proc.
  SPIE} {\bf 9908},  99086C (2016).

\bibitem{carleoetal2016}
{Carleo}, I., {Sanna}, N., {Gratton}, R., {Benatti}, S., {Bonavita}, M.,
  {Oliva}, E., {Origlia}, L., {Desidera}, S., {Claudi}, R., and {Sissa}, E.,
  ``{High precision radial velocities with GIANO spectra},'' {\em Experimental
  Astronomy}~{\bf 41},  351--376 (2016).

\bibitem{gulliksonetal2014aj}
{Gullikson}, K., {Dodson-Robinson}, S., and {Kraus}, A., ``{Correcting for
  Telluric Absorption: Methods, Case Studies, and Release of the TelFit
  Code},'' {\em \aj}~{\bf 148},  53 (Sept. 2014).

\bibitem{kauschetal2015aa}
{Kausch}, W., {Noll}, S., {Smette}, A., {Kimeswenger}, S., {Barden}, M.,
  {Szyszka}, C., {Jones}, A.~M., {Sana}, H., {Horst}, H., and {Kerber}, F.,
  ``{Molecfit: A general tool for telluric absorption correction. II.
  Quantitative evaluation on ESO-VLT/X-Shooterspectra},'' {\em \aap}~{\bf 576},
   A78 (Apr. 2015).

\bibitem{smetteetal2015aa}
{Smette}, A., {Sana}, H., {Noll}, S., {Horst}, H., {Kausch}, W., {Kimeswenger},
  S., {Barden}, M., {Szyszka}, C., {Jones}, A.~M., {Gallenne}, A., {Vinther},
  J., {Ballester}, P., and {Taylor}, J., ``{Molecfit: A general tool for
  telluric absorption correction. I. Method and application to ESO
  instruments},'' {\em \aap}~{\bf 576},  A77 (Apr. 2015).

\bibitem{raineretal2018spie}
{Rainer}, M., {Harutyunyan}, A., {Carleo}, I., {Oliva}, E., {Benatti}, S.,
  {Bignamini}, A., {Claudi}, R., {Gonzalez-Alvarez}, E., {Sanna}, N.,
  {Ghedina}, A., {Micela}, G., {Molinari}, E., {Tozzi}, A., {Baffa}, C.,
  {Baruffolo}, A., {Buchschacher}, N., {Cecconi}, M., {Cosentino}, R.,
  {Falcini}, G., {Fantinel}, D., {Fini}, L., {Galli}, A., {Ghinassi}, F.,
  {Giani}, E., {Gonzalez}, C., {Gonzalez}, M., {Gratton}, R., {Guerra}, J.,
  {Hernandez Diaz}, M., {Hernandez}, N., {Iuzzolino}, M., {Lodi}, M.,
  {Malavolta}, L., {Maldonado}, J., {Origlia}, L., {Perez Ventura}, H.,
  {Puglisi}, A., {Riverol}, C., {Riverol}, L., {San Juan}, J., {Scuderi}, S.,
  {Seeman}, U., {Sozzetti}, A., and {Sozzi}, M., ``{Introducing GOFIO: a DRS
  for the GIANO-B near-infrared spectrograph},'' in [{\em Ground-based and
  Airborne Instrumentation for Astronomy VII}{\nolinebreak\hspace{0.1em}]},
  {Evans}, C.~J., {Simard}, L., and {Takami}, H., eds., {\em Society of
  Photo-Optical Instrumentation Engineers (SPIE) Conference Series} {\bf
  10702},  1070266 (July 2018).

\end{thebibliography}
\bibliographystyle{spiebib} % makes bibtex use spiebib.bst

\end{document}